\documentclass[a4paper,conference]{IEEEtran}
\IEEEoverridecommandlockouts
\usepackage{cite}
\usepackage{amsmath,amssymb,amsfonts}
\usepackage{algorithmic}
\usepackage{graphicx}
\usepackage{textcomp}
\usepackage{xcolor}
\def\BibTeX{{\rm B\kern-.05em{\sc i\kern-.025em b}\kern-.08em
    T\kern-.1667em\lower.7ex\hbox{E}\kern-.125emX}}
\begin{document}

\title{A Hybrid Approach to Enhance Pure Collaborative Filtering based on Content Feature Relationship
}

\author{\IEEEauthorblockN{Mohammad Maghsoudi Mehrabani}
\IEEEauthorblockA{\textit{School of Engineering Science, College of Engineering} \\
\textit{University of Tehran}\\
Tehran, Iran \\
m.maghsoudi.m@ut.ac.ir}
\and
\IEEEauthorblockN{Hamid Mohayeji}
\IEEEauthorblockA{\textit{Department of Computer Engineering} \\
\textit{Sharif University of Technology}\\
Tehran, Iran \\
h.mohayeji92@student.sharif.edu}
\and
\IEEEauthorblockN{Ali Moeini}
\IEEEauthorblockA{\textit{School of Engineering Science, College of Engineering} \\
\textit{University of Tehran}\\
Tehran, Iran \\
moeini@ut.ac.ir}
}

\maketitle

\begin{abstract}
  Recommendation systems get expanding significance because of their applications in both the scholarly community and industry. With the development of extra data sources and methods of extracting new information other than the rating history of clients on items, hybrid recommendation algorithms, in which some methods have usually been combined to improve performance, have become pervasive. In this work, we first introduce a novel method to extract the implicit relationship between content features using a sort of well-known methods from the natural language processing domain, namely Word2Vec. In contrast to the typical use of Word2Vec, we utilize some features of items as words of sentences to produce neural feature embeddings, through which we can calculate the similarity between features. Next, we propose a novel content-based recommendation system which employs the relationship to determine vector representations for items by which the similarity between items can be computed (RELFsim). Our evaluation results demonstrate that it can predict the preference a user would have for a set of items as good as pure collaborative filtering. This content-based algorithm is also embedded in a pure item-based collaborative filtering algorithm to deal with the cold-start problem and enhance its accuracy. Our experiments on a benchmark movie dataset corroborate that the proposed approach improves the accuracy of the system. 
\end{abstract}


\section{Introduction}
There is no doubt that nowadays, the Internet plays a key role in people’s lives, and too much information has become available on the Internet because of which the users have difficulty in finding and choosing the appropriate items among many collections. Therefore, companies and system owners have deployed sophisticated algorithms to provide their customers with recommendation systems (RSs) which help them cope with information overload problems \cite{shi2017long,hwangbo2018recommendation}
.Various studies have been going on around this area of RSs over the last few decades. However, the importance of RSs still remains high due to their applications in different domains such as traveling, news, scientific articles, or advertising\cite{karimi2018news,wang2018content}.
Basically, RSs try to realize the taste of the users according to some available data such as users' ratings on items, purchase history of users, or contextual information on the users or items, and predict the preference of a user for items which have not been seen by the user. Then taking these predictions into account, the most relevant items are suggested to the user. This way, the user is provided with a small proportion of items that are well suited to the user’s taste. 
A necessary set of data to make personalized recommendations is a kind of user feedback on the items which can be explicit like the ratings on the items or implicit like the time a user spends watching details of an item. In addition, other information such as contextual information of items or users can be useful to have a better RS. Nevertheless, it is not easy to obtain contextual information in most cases. 
RSs can be classified into three broad groups: content-based, CF and Hybrid systems. Content-based approaches focus on the properties of items. Indeed, the content of items which are visited by the user is used to recommend other items which have similar content \cite{lops2011content}.
There is a number of reasons why content-based approaches should be used such as not being available or accessible data of other users. CF approaches focus on the user-item interactions. In other words, the ratings on the items as a rating matrix are used to identify what items a user is interested in based on like-minded users \cite{konstan1997grouplens,schafer2007collaborative}.
Collaborative filtering (CF) and content-based, two of the prime and well-known approaches, work based on user-item interactions (ratings on items) and contextual information respectively \cite{balabanovic1997fab}. Both approaches have their own advantages and disadvantages \cite{pazzani2007content,balabanovic1997fab} as a result of which hybrid approaches, which are generally created by combining other methods, have become popular which attempt to enjoy advantages of both aforementioned approaches and overcome their drawbacks \cite{burke2002hybrid,balabanovic1997fab}. For instance, CF algorithms have been successfully used in most situations because they can work with only interaction data regardless of the unavailability of contextual information, but they have some issues one of the most important of which is cold-start which indicates that the amount of ratings for new items or users is not adequate to prepare appropriate recommendations \cite{schein2002methods}. Word embedding methods - a set of techniques in natural language processing (NLP) - can also be used to map words from text information of items to vectors of real numbers as word representations by which identifying similar textual information of items is possible \cite{musto2016learning}. Word2Vec is a set of related models architectures used to compute vector representations of words using two-layer neural networks \cite{mikolov2013distributed,mikolov2013efficient}. One of the model architectures is the skip-gram model which is an efficient method for learning high-quality distributed vector representations that capture a large number of precise syntactic and semantic word relationships according to what Mikolov et al. introduced on their paper \cite{mikolov2013distributed,mikolov2013efficient}.
In this paper, we present a new concept of the relationship between content features and propose a novel content-based algorithm based on the relationship, and then we use the proposed content-based to design a hybrid RS. The hybrid one aims to deal with the cold-start problem in a pure CF RS to improve the accuracy of the system.
%\begin{itemize}
% \item Introducing the concept of content feature relationship and an innovative method for extracting such relationship; 
% \item Introducing a new approach for dealing with the cold start problem (when there are new item items to be recommended) in item-based CF algorithm; 
% \item Proposing an improved content-based RS, which outperforms the state-of-the-art top-N recommendation methods.
%\end{itemize}
   
\section{Related Work}   
   This section presents a brief review of literature which is related to dealing with the cold-start problem in CF using content-based methods based on new ways of computing content similarity.
Some hybrid methods have been proposed which employ different information sources to enhance CF algorithms or employ word embedding techniques to gain advantages from textual information. Melville et al. put forward content-boosted CF which exploits textual information of movies (e.g. title, cast, etc.) as features to enhance the rating matrix with which CF can work better \cite{melville2002content}.
Mobasher et al. presented SimComb which brings structured semantic knowledge about items (ontologies) into play to cope with cases in which little or no rating is available for new items as a result of which the accuracy is improved \cite{mobasher2003semantically}.
Gunawardana and Meek suggested unified Boltzmann machines (probabilistic models) which encode collaborative and content information as features to learn weights that reflect the importance of different pairwise interactions \cite{gunawardana2009unified}.
Lin et al. described a method that considers the nascent information culled from Twitter to provide relevant recommendations in cold-start situations \cite{lin2013addressing}. 
Kouki et al. performed HyPER (a probabilistic model) which is a general hybrid framework, and it is able to combine multiple information types from different sources and modeling techniques into a single unified model to enhance the performance of the RS \cite{kouki2015hyper}.
Aslanian et al. introduced hybrid RS algorithms based on content feature relationship which is extracted from the rating matrix using a mathematical formulation \cite{jalili2016hybrid}.
Wei et al. proposed two models which extract content features of the items using deep neural networks which are taken into the prediction of ratings for the cold-start items \cite{wei2017collaborative}.
Nilashi et al. developed a new hybrid RS using combinations of dimensionality reduction and ontology techniques to find the most similar items and users in order to solve sparsity and scalability problems \cite{nilashi2018recommender}.
 Musto et al. employed word embedding techniques to learn a low- dimensional vector space word representation from textual information of Wikipedia to represent both items and user profiles in a content-based recommendation \cite{musto2015word}.
Musto et al. developed a content-based recommendation framework using semantic vectors for movies that are extracted from Wikipedia using word embedding techniques \cite{musto2016learning}. 
Ozsoy applied Word2Vec to RSs domain to capture the correlation between venues and to recommend new venues to users \cite{ozsoy2016word}.
   
\section{Overview of the Techniques}

\subsection{Pure item-based collaborative filtering}
Item-based CF focuses on the similarity of the user ratings for two items \cite{leskovec2014mining}. Indeed, two items are similar if users, who have rated both, have given both items similar ratings. To compute the similarity between items we use cosine similarity as a similarity measure which is defined bellow \cite{leskovec2014mining}:
\begin{equation} \label{eq:cosine_sim}
\operatorname {sim} (i,j)=\cos({\vec {i}},{\vec {j}})={\frac {{\vec {i}}\cdot {\vec {j}}}{||{\vec {i}}||\times ||{\vec {j}}||}}={\frac {\sum \limits _{u\in U_{ij}}R_{u,i}R_{u,j}}{{\sqrt {\sum \limits _{u\in U_{ij}}R_{u,i}^{2}}}{\sqrt {\sum \limits _{u\in U_{ij}}R_{u,j}^{2}}}}}
\end{equation}
where $U_{ij}$ is a subset of users who rated both items (i,j) , \textit{R} is the rating matrix in which columns represent items, rows represent users and $R_{u,i}$ represents the rating that user \textit{u} given to item \textit{i}. In addition, we define  $I= \{i_{1}, i_{2},i_{3},...,i_{M} \} $ as a set of items, $U=\{u_1,u_2,...,u_N\} $ as a set of users.To predict $R_{u,i}$, first the similarities between item \textit{i} and other items are measured, and then \textit{k} items that have the highest similarity with item \textit{i} are selected as the neighborhood. Finally, the value of $R_{u,i}$ is computed from weighted average of the neighbors' ratings as follow:
\begin{equation} \label{eq:weighted_mean}
{\displaystyle \operatorname {R_{u,i}}=
{\frac {\sum \limits _{j \in k\_Neighbors(i)} sim(i,j).(R_{u,j}-\bar{R_{j}}) }
{\sum \limits _{j \in k\_Neighbors(i)}sim(i,j)}
} + \bar{R_{i}}}
\end{equation}
where ${\bar {R_{i}}} $ is the average rating given to item $i$.
\subsection{Word2Vec}
\begin{figure}[htbp]
\centerline{\includegraphics[width=.45\textwidth]{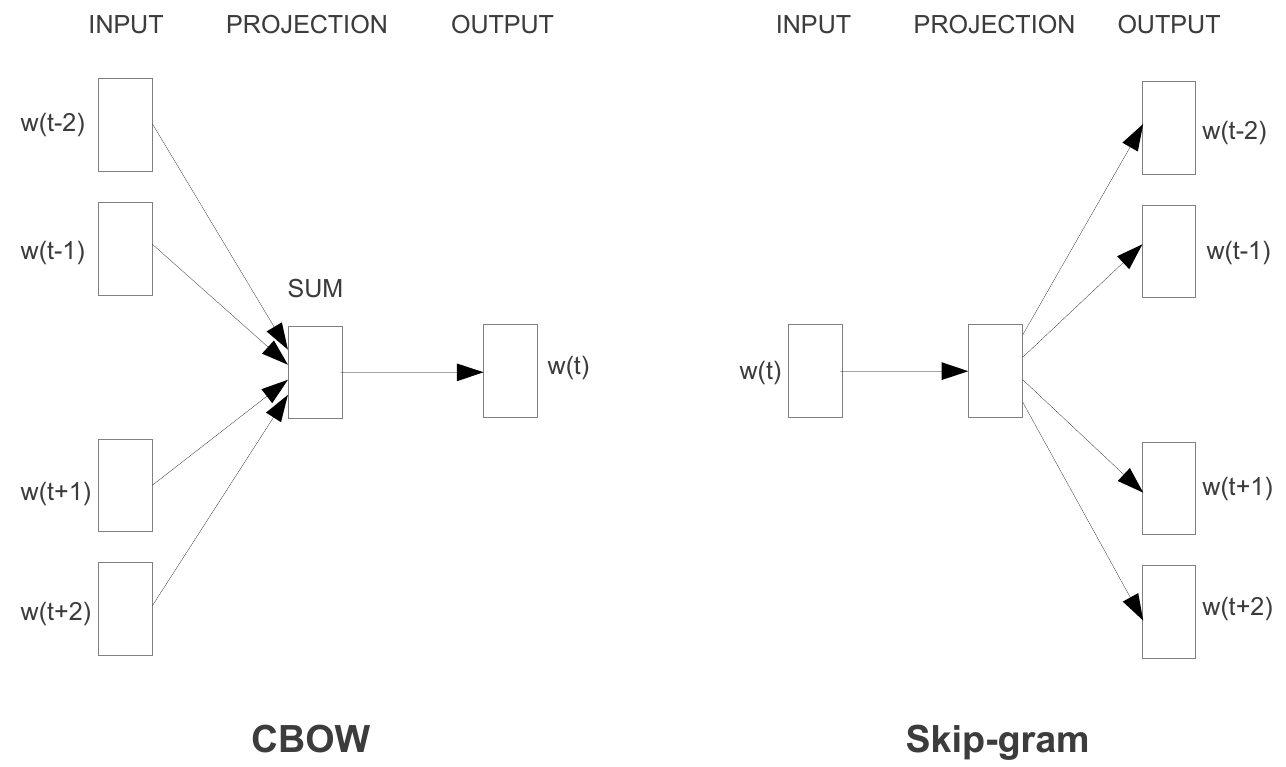}}
\caption{The CBOW architecture predicts the current word based on the context, and the Skip-gram predicts surrounding words given the current word\cite{mikolov2013efficient}.}
\label{word2vec}
\end{figure}

Word2Vec is a set of related models architectures used to compete vector representations of words using two-layer neural networks\cite{mikolov2013distributed,mikolov2013efficient}. Briefly, there are two model architectures for learning distributed representations of words, namely continuous bag-of-words (CBOW) and continuous skip-gram\cite{mikolov2013efficient}. CBOW uses continuous distributed representation of the context to predict the currunt word, the second architecture is similar to CBOW, but tries to predict the context (words around the currunt one) using the currunt word \cite{mikolov2013efficient} (Figure \ref{word2vec}).
The input of Word2Vec is a set of sentences, each of which is a sequence of words, and Skip-gram tries to capture the sequential nature of the sentences by considering some words surround each word. After training the model, the hidden layer of the model is a set of vectors, which are vector representations of the input words that are learned. According to what the network does, two dissimilar words that have similar contexts in the sentences provided with close vectors. Therefore, one can compute the similarity between two words by computing the cosine similarity between their vectors. 

\section{Proposed methods based on content feature relationship}
\subsection{Extracting the content feature relationship}
\begin{figure}[htbp]
\centerline{\includegraphics[width=.49\textwidth]{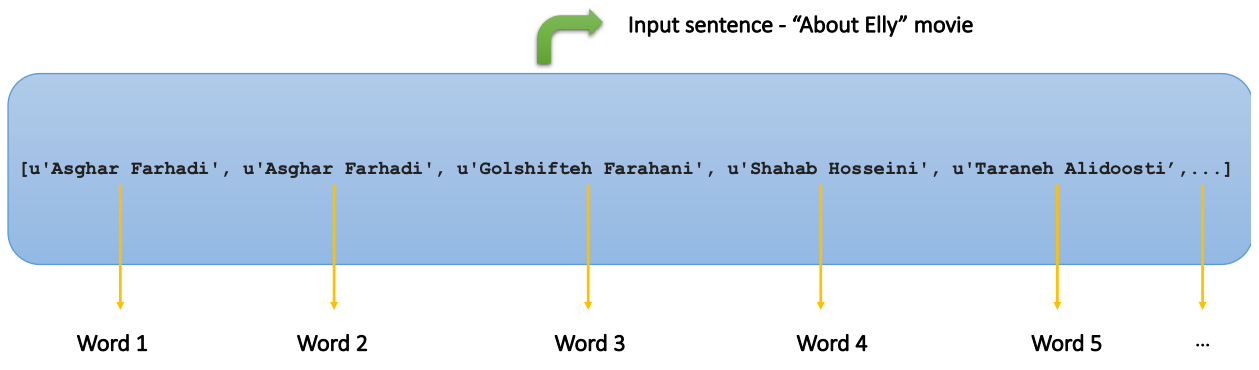}}
\caption{An example of the sequence of words in the sentence for a movie: 'word 1 : the director' + 'word 2 : the screenwriter' + 'word 3 : the first actor' + 'word 3 : the second actor' + ... . }
\label{word2vec_input}
\end{figure}

First, we opt for the most effective features and list them in order of importance. As the dataset used here is a movie dataset, we make a list of directors, screenwriters and the first twelve members of the cast in order of appearance for each movie. We consider this list of a movie to be a sentence for that movie (Figure \ref{word2vec_input}). Then we use a list of the sentences as the input of the skip-gram model. After the training step, we have the vector representations for directors, screenwriters, and actors.
Our justification for the approach is that two actors who have played several roles in similar contexts with the same actors, directors or screenwriters, are similar. This relationship is what can be learned using skip-gram architecture. Moreover, the closer the positions of two actors, who play roles in the same movie, are in the sentence of the movie, the more similar their vectors will be. Finally, we can compute the cosine similarity between vector representations of each feature to measure the relationship between them
The hyperparameters of the skip-gram architecture have effects on the accuracy of the RS. We set "Window Size", "Vector dimension", "Negative", "Min-count" and "epoch" 8, 150, 25, 1 and 20 respectively.

\subsection{A content-based predictor using the relationship}
To exploit the extracted content feature relationship, there are several ways, but we simply take the average of word vectors of each movie to obtain a vector representation for that movie. Averaging the embeddings of words in a sentence tested as a successful and efficient way of obtaining sentence embeddings\cite{kenter-etal-2016-siamese}. In this way, we can compute the cosine similarity between movies as a similarity considering the relationship - called RELFsim. Now we implement a content-based predictor using a pure CF algorithm. In fact, we exploit RELFsim between movies instead of computing the cosine similarity between the ratings of movies in a CF algorithm. Thus, we have a content-based predictor that does not have any problem with new items. In comparison with pure content-based filtering, this one can perform a similarity between two movies that have no actors in common, but their actors have appeared in the same movies in the past. 

\subsection{Combining the content-based predictor with pure collaborative filtering to deal with the cold-start problem}
As our proposed content-based algorithm employs the pure CF algorithm to predict the ratings, we can combine them easily. Whenever new items are added or few ratings are submitted for some items, the CF algorithm is not able to predict ratings for those items, and pure CF cannot recommend these kinds of items. Therefore, when CF tries to predict the rating of item \textit{i} for user \textit{u}, in the step that similarity between item \textit{i} and other items, which are rated by user \textit{u}, should be computed, we check whether there are enough ratings for items or not. If there were no rating or few ratings, it would bring RELFsim into play. As a result, the RS can predict the ratings for items with few ratings or no ratings as well as for others.

\section{Experiments}
\subsection{Dataset}
Our experiments have been performed on MovieLens 1M Dataset, a stable benchmark dataset, contains 1,000,209 anonymous ratings of approximately 3,900 movies made by 6,040 MovieLens users who joined MovieLens in 2000 \cite{MovieLens_dataset}.To have cast and crew for movies, we use TMDb information and use the Links file of MovieLens, which contains the IDs of movies on other sources, to merge them. There were a few duplicated rows and invalid values for IDs filed and some movies without information that cleaned. After cleaning the data, 995138 ratings of 3746 movies remained. We use only directors, screenwriters and first twelve actors of each movie if they exist. In the model 22669 unique features used that means 22669 vectors created. 
%To evaluate the results we divide the data into two Trainset and Testset that are 80\% and 20\% of the data respectively.
\subsection{The Content-based predictor and the Hybrid approach}
\begin{figure}[htbp]
\centerline{\includegraphics[width=.49\textwidth]{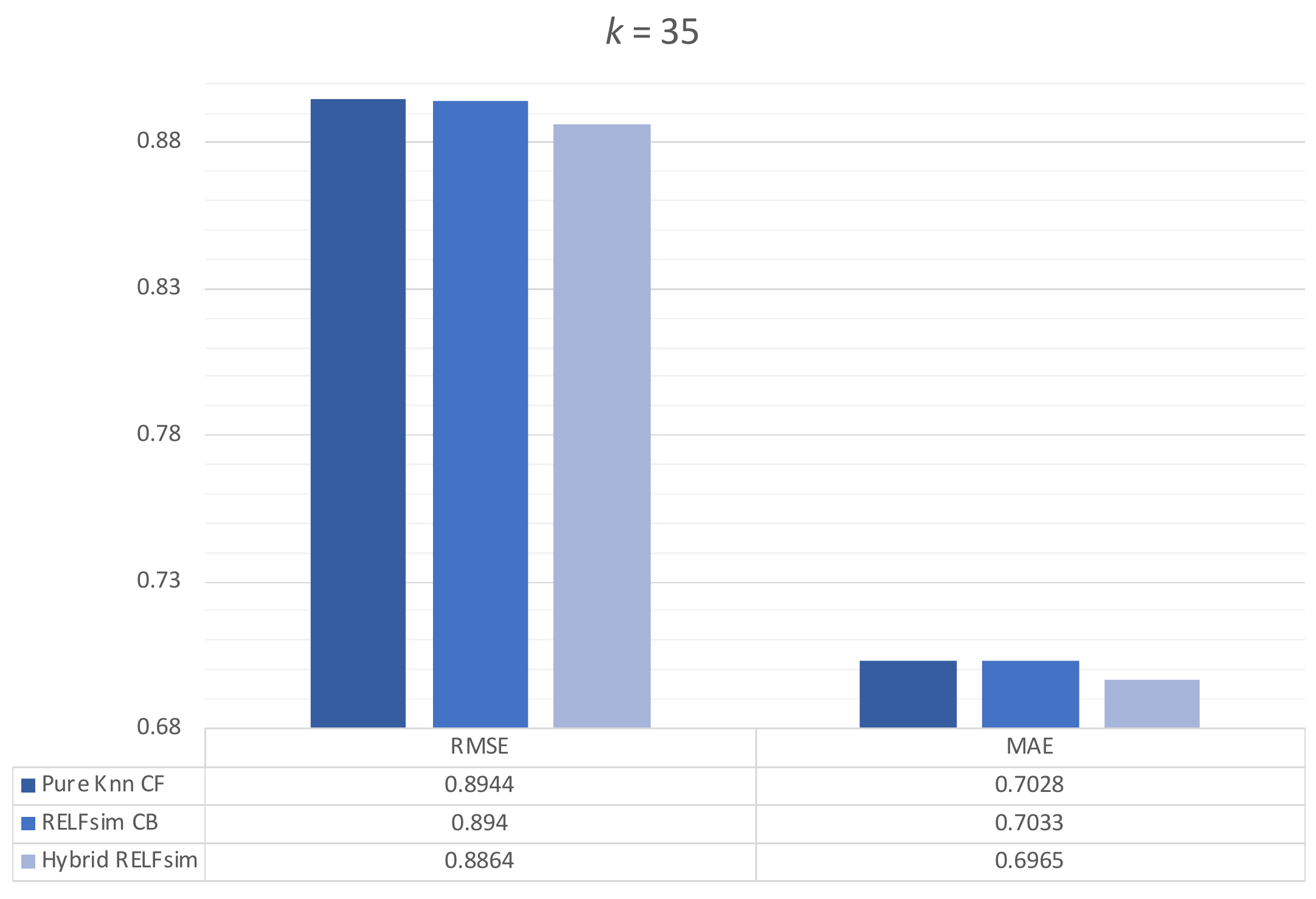}}
\caption{Root Mean Square Error (RMSE) and Mean Absolute Error (MAE) using three methods include the proposed content-based predictor (RELFsim CB), the proposed hybrid RS (Hybrid RELFsim and the pure CF (Pure Knn CF).}
\label{rmse_ame}
\end{figure}

In this section, we compare the accuracy of the proposed content-based predictor, the proposed Hybrid RS, and pure CF. To measure the accuracy of the prediction task, we use two well-known metrics Root Mean Square Error (RMSE) and Mean Absolute Error (MAE). They show how close RSs predict the ratings for users and items in the test set. Once we used five-fold cross-validation to cover all the data as a test set and $k = 35$ as the number of selected neighbors. Then to evaluate the effectiveness of the value of \textit{K}, we divided the data into two Trainset and Testset that are 80\% and 20\% of the data respectively and ran the algorithms with the diverse values of \textit{k}. We did not modify the data to form cold-start items and there are a few items in the cold-start situation. As it is obvious, the more cold-start items were there, the more hybrid CF would enhance the accuracy of the pure CF. As it is seen in figure \ref{rmse_ame}, the proposed content-based works as good as the pure CF and the combination of them can enhance the accuracy and be effective to deal with the cold-start problem.

\begin{figure}[htbp]
\centerline{\includegraphics[width=.49\textwidth]{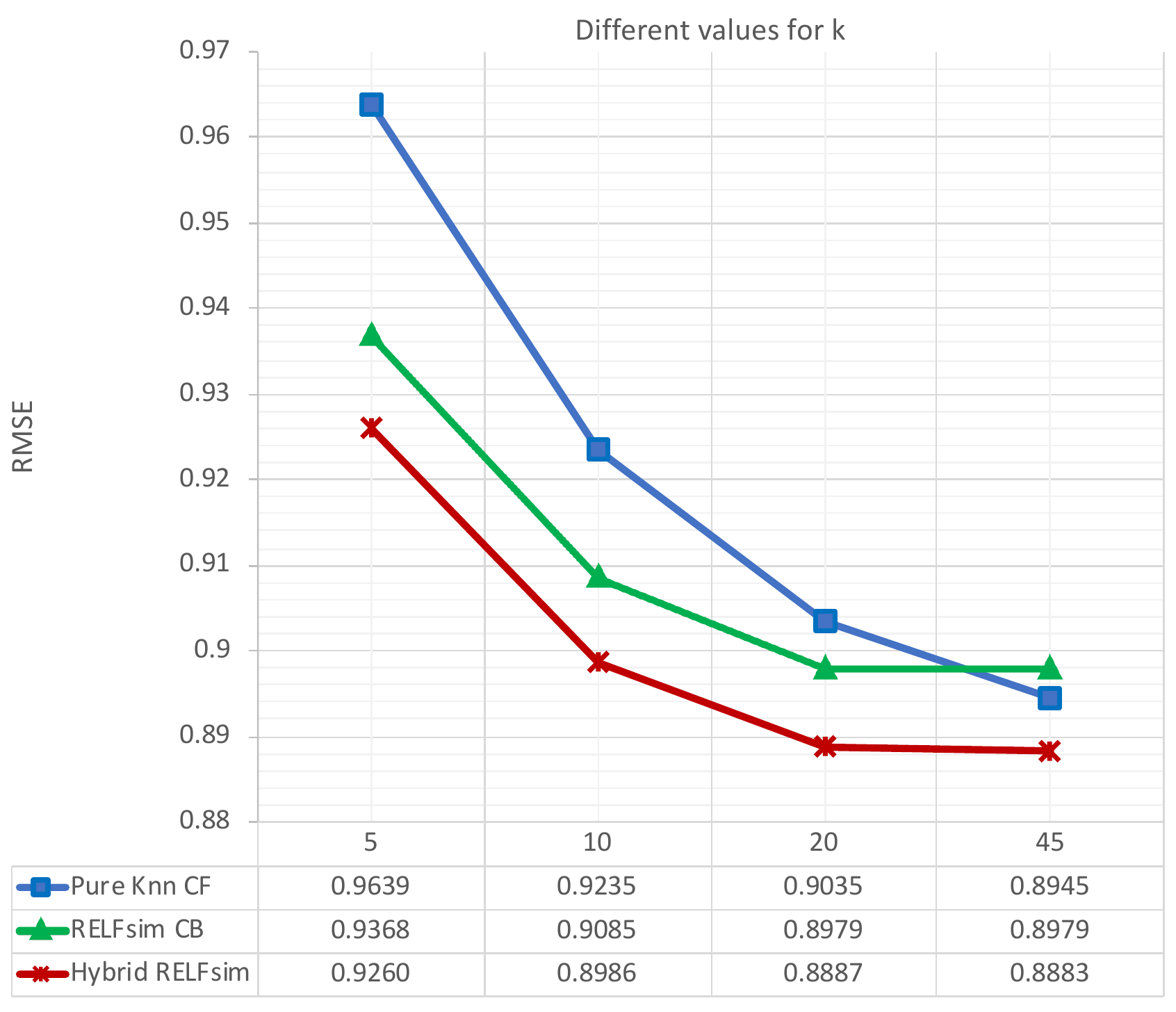}}
\caption{The effect of parameter \textit{k} on RMSE for three methods including the proposed content-based predictor (RELFsim CB), the proposed hybrid RS (Hybrid RELFsim and the pure CF (Pure Knn CF).}
\label{rmse_k}
\end{figure}

\begin{figure}[htbp]
\centerline{\includegraphics[width=.49\textwidth]{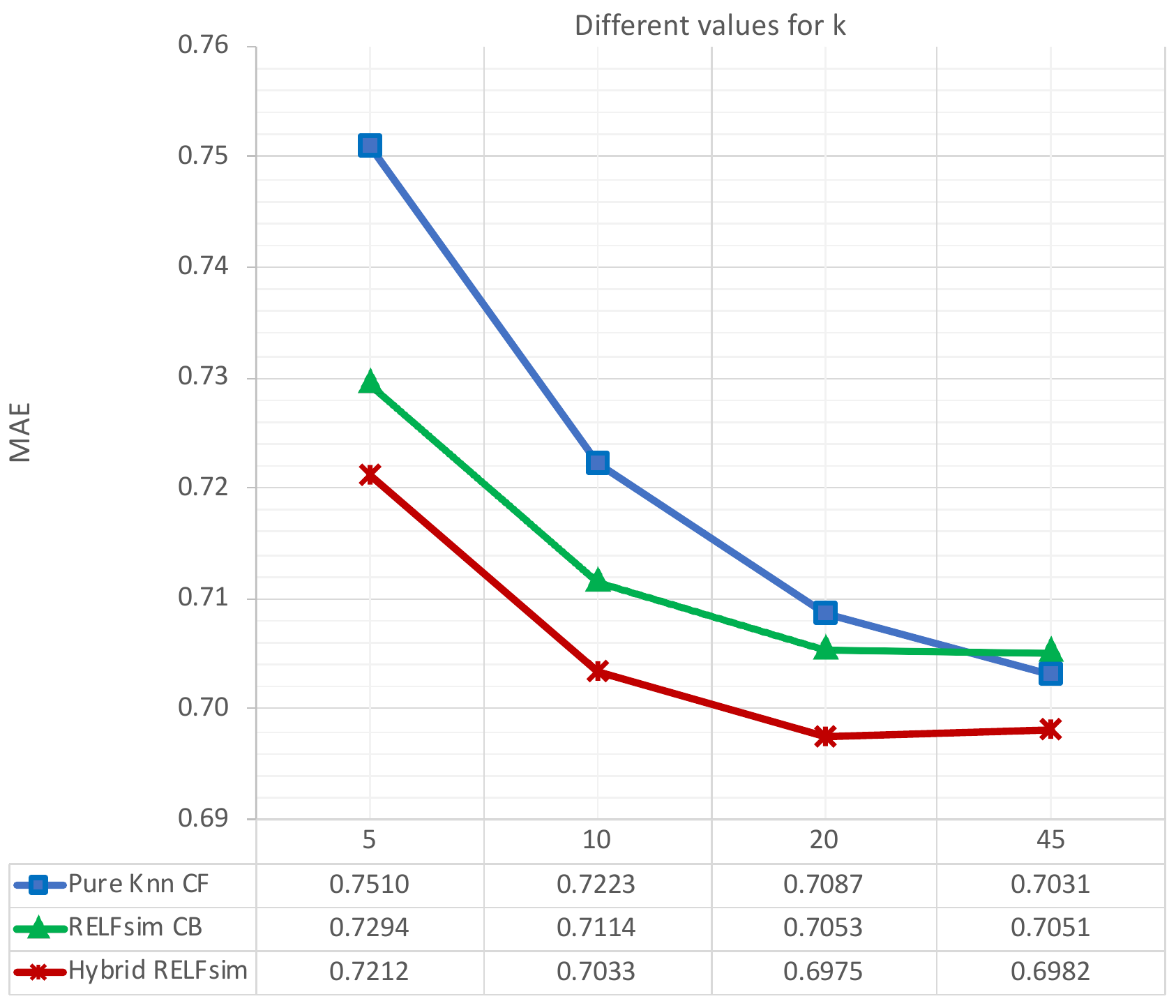}}
\caption{The effect of parameter \textit{k} on MAE for three methods including the proposed content-based predictor (RELFsim CB), the proposed hybrid RS (Hybrid RELFsim and the pure CF (Pure Knn CF).}
\label{mae_k}
\end{figure}

Figures \ref{rmse_k} and \ref{mae_k} show the influence of parameter k on RMSE and MAE based on our practical experience. From the empirical evidence, it seems that pure CF works very well with sufficient data, but suffers from a lack of enough data in cold-start condition and the combination of pure CF and RELFsim CB can alleviate the problem and result in a better prediction error.

\section{Conclusion}
In this paper, we proposed an innovative learning-based approach to learn and extract the content feature relationship according to the history of emerging features in similar contexts. This relationship was learned using Word2Vec architecture particularly the skip-gram model \cite{mikolov2013efficient} that typically used to produce word embeddings. In this approach, we utilized the relationship to propose a content-based predictor method based on this relationship of content features that works as good as pure CF method and combined it with a pure item-based CF method to advance a hybrid RS that could deal with the cold-start problem to some extent. One of the benefits of this method is the capability of applying the least content features to have a comparable content-based capture some relationship between two items that have even no feature in common. Our experiments attest to the improvement of the accuracy of pure CF.
As future work, we will work on complex methods of exploiting the relationship to produce vectors for items because in this work we adopted a simple way to create vectors for items taking the average of their features vectors to realize that the relationship could be useful. Besides, we had an uncomplicated mixture of two algorithms to perform a hybrid RS. However, there are various methods can be involved to create hybrid RSs like strategies in content-boosted CF \cite{melville2002content} and semantically enhanced CF \cite{mobasher2003semantically} on which we are going to study.

\section*{Acknowledgment}

The authors would like to acknowledge the financial support from the University of Tehran for this research.

\bibliographystyle{IEEEtran}
\bibliography{IEEEabrv,Reference}

\end{document}